\documentclass[12pt]{iopart}
\usepackage{xcolor}
\usepackage{url}
\usepackage{graphicx}
\expandafter\let\csname equation*\endcsname\relax
\expandafter\let\csname endequation*\endcsname\relax
\usepackage{amsmath}
\usepackage{hyperref}

\begin{document}

\title{CP2K on the road to exascale}

\author{Thomas D. Kühne$^{\dagger,\mathsection,*}$, Christian Plessl$^{\ddagger,\mathsection}$, Robert Schade$^{\ddagger,\mathsection}$, and Ole Schütt$^\mathparagraph$}

\address{$^\dagger$ Dynamics of Condensed Matter and Center for Sustainable Systems Design, Chair of Theoretical Chemistry, Paderborn University, D-33098 Germany \\
$^\ddagger$Department of Computer Science, Paderborn University, D-33098 Germany\\
$^\mathsection$ Paderborn Center for Parallel Computing, Paderborn University, D-33098 Germany
$^\mathparagraph$CP2K Foundation, CH-8006 Zurich Switzerland}

\ead{tkuehne@cp2k.org} 
\vspace{10pt}

\begin{abstract}

The CP2K program package, which can be considered as the swiss army knife of atomistic simulations, is presented with a special emphasis on \textit{ab-initio} molecular dynamics using the second-generation Car-Parrinello method. After outlining current and near-term development efforts with regards to massively parallel low-scaling post-Hartree-Fock and eigenvalue solvers, novel approaches on how we plan to take full advantage of future low-precision hardware architectures are introduced. Our focus here is on combining our submatrix method with the approximate computing paradigm to address the immanent exascale era.
\end{abstract}

%
%
%
%
%

\section{Background and Current Status}

The open-source simulation package CP2K is an extensive quantum chemistry and condensed matter physics program that comprises a large variety of different theoretical methods and computational approaches to conduct most diverse atomistic simulations for large-scale condensed-phase systems, such as liquids, solids, nanomaterials and molecular structures, to name just a few. 
All of this is made possible from the outset by the design of highly efficient algorithms with a focus on excellent parallel scalability and suitability for novel high-performance computing architectures, as demonstrated by its \textsc{Quickstep} electronic structure module. The latter is based on the Gaussian and plane wave (GPW) approach and its all-electron variant, the  Gaussian-augmented plane wave (GAPW) method, which allows for a particularly efficient treatment of large-scale orbital-free and Kohn–Sham density functional theory (DFT) calculations, as well as wavefunction-based correlation methods, such as Hartree-Fock (HF), second-order M{\o}ller–Plesset perturbation theory (MP2), random phase approximation (RPA), hybrid and double-hybrid DFT and the GW approximation, all with arbitrary boundary conditions \cite{CP2K}.

However, the great appeal of CP2K lies in the possibility to combine all available total energy and force methods with a wide selection of sampling techniques, such as Monte Carlo, Ehrenfest/real-time dynamics and most importantly molecular dynamics, as well as advanced free-energy and rare-event sampling schemes, to enable realistic simulations at finite-temperature beside conventional static calculations. On the one hand, this necessitates the general availability of analytic gradients in particular for periodic boundary conditions, in order to permit the efficient calculation of nuclear forces. 
On the other hand, minimum time to solution is essential to allow for an extensive sampling via the techniques listed above. In that respect, a unique selling point of CP2K is the implementation of the second-generation Car-Parrinello method, which allows to routinely conduct nanosecond long DFT-based \textit{ab-initio} molecular dynamics (AIMD) simulations with thousands of atoms. The superior efficiency of this approach originates from the design of an improved coupled electron-ion dynamics that keeps the electrons very close to their instantaneous ground-state using just one preconditioned gradient calculation per AIMD step, which can thus be thought of as an electronic force to propagate the electronic degrees of freedom in dimensionless time \cite{CP2G}.

\section{Development Priorities}

Beside the implementation of sophisticated spectroscopic properties \cite{EXAFS1, EXAFS2}, which are either based on density functional perturbation theory or time-dependent DFT within the Tamm-Dancoff approximation \cite{CP2K,TDDFT}, the current development priorities are focused mainly on devising novel low-scaling post-HF methods including the implementation of analytic nuclear gradients for arbitrary boundary conditions. As already indicated above, particular emphasize is on HF and MP2 methods, which are a prerequisite for simulations based on RPA, hybrid and double-hybrid DFT schemes, as well as GW. The four-center two-electron repulsion integrals (ERI), which in Mulliken notation reads as 
\begin{equation}
(\mu \nu | \lambda \sigma) = \int d\mathbf{r}_1 \int d\mathbf{r}_2 \, \phi_{\mu}^*(\mathbf{r}_1) \phi_{\nu}(\mathbf{r}_1) \, \frac{1}{|\mathbf{r}_1 - \mathbf{r}_2|} \, \phi_{\lambda}^*(\mathbf{r}_2) \phi_{\sigma}(\mathbf{r}_2) 
\end{equation}
are of central importance for all wavefunction-based post-HF methods.  
In addition to well established integral screening techniques based on the Schwarz inequality
\begin{equation}
  |(\mu \nu | \lambda \sigma)| \leq |(\mu \nu | \mu \nu)|^{1/2} \, |(\lambda \sigma | \lambda \sigma)|^{1/2} 
\end{equation}
to reduce the scaling from $\mathcal{O}(N^4)$ to $\mathcal{O}(N^2)$, a similar density matrix screening can also be employed to eventually sustain linear scaling with respect to the system size $N$. In the latter, the largest density matrix element $P_{\max} = \max \{ |P_{\mu \lambda}|, |P_{\mu \sigma}|, |P_{\nu \lambda}|, |P_{\nu \sigma}|\}$ is used to screen elements smaller than $\epsilon_{\text{Schwarz}}$ via  \begin{equation} 
P_{\max} \times |(\mu \nu | \mu \nu)|^{1/2} \, |(\lambda \sigma | \lambda \sigma)|^{1/2} \leq \epsilon_{\text{Schwarz}}, 
\end{equation}
where $P_{\max}$ is either the density matrix from the previous self-consistent field iteration, or from a converged semi-local DFT calculation, but ideally the propagated density matrix via second-generation Car-Parrinello AIMD of the previous timestep \cite{ADMM}. 
Yet, at the core of all implemented post-HF approaches are either the auxiliary density matrix method (ADMM) \cite{ADMM}, or the resolution of identity (RI) approach \cite{RI-MP2-RPA2}. The key ingredient of the former is the use of an auxiliary density matrix $\mathbf{\hat{P}}$, which approximates the original density matrix $\mathbf{P}$, but is substantially easier to compute due to being smaller in size, or more rapidly decaying than $\mathbf{P}$. For the case of computing the Hartree-Fock exchange (HFX), the exact energy $E_{\text{X}}^{\text{HFX}}[\mathbf{P}]$ is replaced by the computationally superior expression $E_{\text{X}}^{\text{HFX}}[\mathbf{\hat{P}}]$, whereas the difference between the two terms is corrected approximately at the semi-local DFT level. Hence, 
\begin{eqnarray}
E_{\text{X}}^{\text{HFX}}[\mathbf{P}] &=& E_{\text{X}}^{\text{HFX}}[\mathbf{\hat{P}}] + \left( E_{\text{X}}^{\text{HFX}}[\mathbf{P}] - E_{\text{X}}^{\text{HFX}}[\mathbf{\hat{P}}] \right) \nonumber \\
&\approx& E_{\text{X}}^{\text{HFX}}[\mathbf{\hat{P}}] + \left( E_{\text{X}}^{\text{DFT}}[\mathbf{P}] - E_{\text{X}}^{\text{DFT}}[\mathbf{\hat{P}}] \right), 
\end{eqnarray}
where $E_{\text{X}}^{\text{DFT}}[\mathbf{P}]$ and $E_{\text{X}}^{\text{DFT}}[\mathbf{\hat{P}}]$ are the DFT exchange energies for the primary and auxiliary density matrices, respectively. The RI approximation, however, allows to substitute the computationally demanding four-center ERIs by just 2- and 3-center integrals by factorizing them via 
\begin{eqnarray}
  (i a | j b)_{\text{RI}} = \sum_{PQ} \, (i a | P) (P | Q)^{-1} (Q | j b), 
\end{eqnarray}
where $(P | Q)^{-1}$ is the inverse of the Coulomb metric over auxiliary Gaussian basis functions, i.e. 
\begin{eqnarray}
  (P | Q) = \int d\mathbf{r}_1 \int d\mathbf{r}_2 \, \phi_P(\mathbf{r}_1) \, \frac{1}{|\mathbf{r}_1 - \mathbf{r}_2|} \, \phi_Q(\mathbf{r}_2).
\end{eqnarray}
Since the latter is a positive definite matrix, its inverse can be efficiently obtained by means of the Cholesky decomposition
\begin{eqnarray}
  (P | Q) = \sum_R \, L_{PR} L_{RQ}^T
\end{eqnarray}
followed by an inversion of the triangular matrix $\mathbf{L}$, i.e.
\begin{eqnarray}
  (P | Q)^{-1} = \sum_R \, L_{PR}^{-T} L_{RQ}^{-1}. 
\end{eqnarray}
In this way, the factorization of the integrals $(i a | j b)$ can be written in compact form as 
\begin{eqnarray}
  (i a | j b)_{\text{RI}} = \sum_P \, B_P^{ia} B_P^{jb}, 
\end{eqnarray}
where 
\begin{eqnarray}
  B_P^{ia} = \sum_R \, (i a | R) L_{PR}^{-1}. 
\end{eqnarray}
Therein, the three-center integrals $(i a | R)$ are computed starting from integrals over atomic orbitals $(\mu \nu | R)$, so that the final expression for the elements $B_P^{ia}$ reads as 
\begin{eqnarray}
  (i a | P) = \sum_{\nu} \, C_{\nu a} \sum_{\mu} \, C_{\mu i} \sum_R \, (\mu \nu | R) L_{PR}^{-1}, 
\end{eqnarray}
where $\mathbf{C}$ is the molecular orbital coefficient matrix, i.e. $\mathbf{CC}^T = \mathbf{P}$. 

Another major development direction is the design of novel linear-scaling algorithms \cite{SchadeDFT}. Within CP2K, the relation 
\begin{eqnarray}
\text{sign}
\begin{pmatrix}
\mathbf{0} & \mathbf{A}\\
\mathbf{I} & \mathbf{0}
\end{pmatrix}
= \text{sign}
\begin{pmatrix}
\mathbf{0} & \mathbf{A}^{1/2}\\
\mathbf{A}^{-1/2} & \mathbf{0}
\end{pmatrix}
\end{eqnarray}
is employed together with various iterative methods to compute the matrix sign function 
\begin{eqnarray}
\text{sign} \left( \mathbf{A} \right) = \mathbf{A} \left( \mathbf{A}^2 \right)^{-1/2}
\end{eqnarray}
to yield the inverses and (inverse) square roots of large sparse matrices $\mathbf{A}$ with a computational effort that scales just linearly with system size. Most importantly, however, the sign function can also be used for the purification of the Kohn-Sham matrix $\mathbf{H}_{\text{KS}}$ into $\mathbf{P}$ by the use of
\begin{eqnarray}
  \mathbf{P} = \frac{1}{2} \left( \mathbf{I} - \text{sign}\left( \mathbf{S}^{-1} \mathbf{H}_{\text{KS}} - \mu \mathbf{I} \right) \right) \mathbf{S}^{-1}.
\label{eq:purification}
\end{eqnarray}
\begin{figure}
    \centering
    \includegraphics[width=\textwidth]{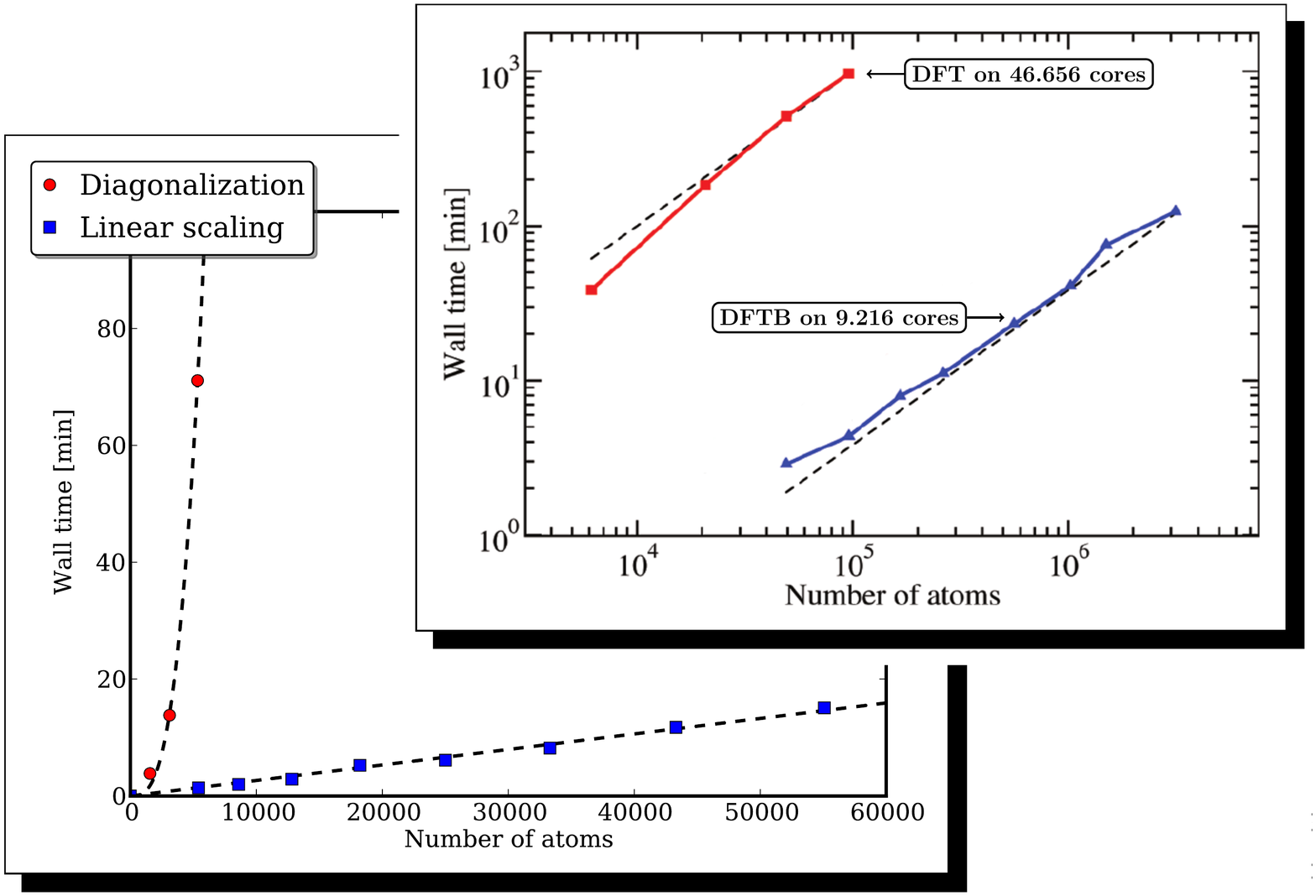}
    \caption{Total walltime for a single AIMD step for bulk water at ambient conditions as a function of system size on a Cray XT5.}
    \label{fig:signmehtod}
\end{figure}
The superior efficiency and linear-scaling potential of this approach has been demonstrated at the DFT level using a double-$\zeta$ valence polarization basis set, for up to a million of atoms in a massively parallel implementation, as shown in Fig.~\ref{fig:signmehtod}.

\section{Meeting the Exascale Challenges}


Standing at the brink of the exascale era, we can see that fundamental disruption in programming models and processor technology that were predicted a decade ago did not happen. From the developer's point of view,  three trends originating in the petascale area are continued and emphasized: 1) the need of massive parallelism in terms of node and thread counts, 2) an even more widespread use of GPUs, and 3) the availability of mixed-precision matrix or tensor operation hardware accelerators. 


CP2K is fundamentally well-suited to scale to immense levels of parallelism because it was designed as a massively parallel MPI application right from the outset. With the advent of multi-core processors, OpenMP directives were added to important loops, while leaving the underlying data layout unchanged to support multi-threading and hybrid parallelism.
%
With that many concurrent threads the data has to be partitioned to prevent bottlenecks from reductions or atomic access collisions.
The new library for Distributed Block-sparse Matrices (DBM) is a first step in this direction. It uses a fixed assignment of matrix block rows to threads, which eliminates the need for synchronization from most operations. 
We plan to refactor other primitives in the same way, in particular the grid data structures that power methods like GPW have a lot to gain from per-thread partitioning.
%
%


The modular structure and use of modern Fortran 2008 allowed the addition of GPU support at the level of GPU-accelerated libraries (e.g. DBM, COSMA, SpFFT, SPLA, grid, pw and Sirius), some of which were also spun off from CP2K as stand-alone libraries such as e.g. DBCSR.
The challenge for the exascale era is the evolution and increasing diversity of GPU architectures. In the early days we mostly struggled with finding the sweet spot within the tight constrains set by small register files and scarce shared memory. 
In present systems, 
the bottleneck has now shifted to the PCI bus, where we are often limited by host-to-device communication. The way forward is to use GPU-aware MPI, which we are currently adding to DBM and will later roll our to other parts of the code.
%
%
The status and results of these efforts can be tracked on the CP2K GPU dashboard at \url{https://www.cp2k.org/gpu}.


Lastly, we expect that the matrix and tensor processing units 
will have a profound and long-lasting effect on method and code development. These computing elements can achieve one to two order of magnitude more FLOPs for dense linear algebra operations in reduced or mixed precision, e.g. FP16 operations with FP32 accumulation. To exploit this potential, it is key to develop methods that heavily rely on dense linear algebra on medium sized local matrices, 
whose numerical inaccuracies due to mixed- and low-precision arithmetic can be rigorously compensated by the design of a modified Langevin-type equation \cite{CP2G}.

\begin{figure}
    \centering
    \includegraphics[width=\textwidth]{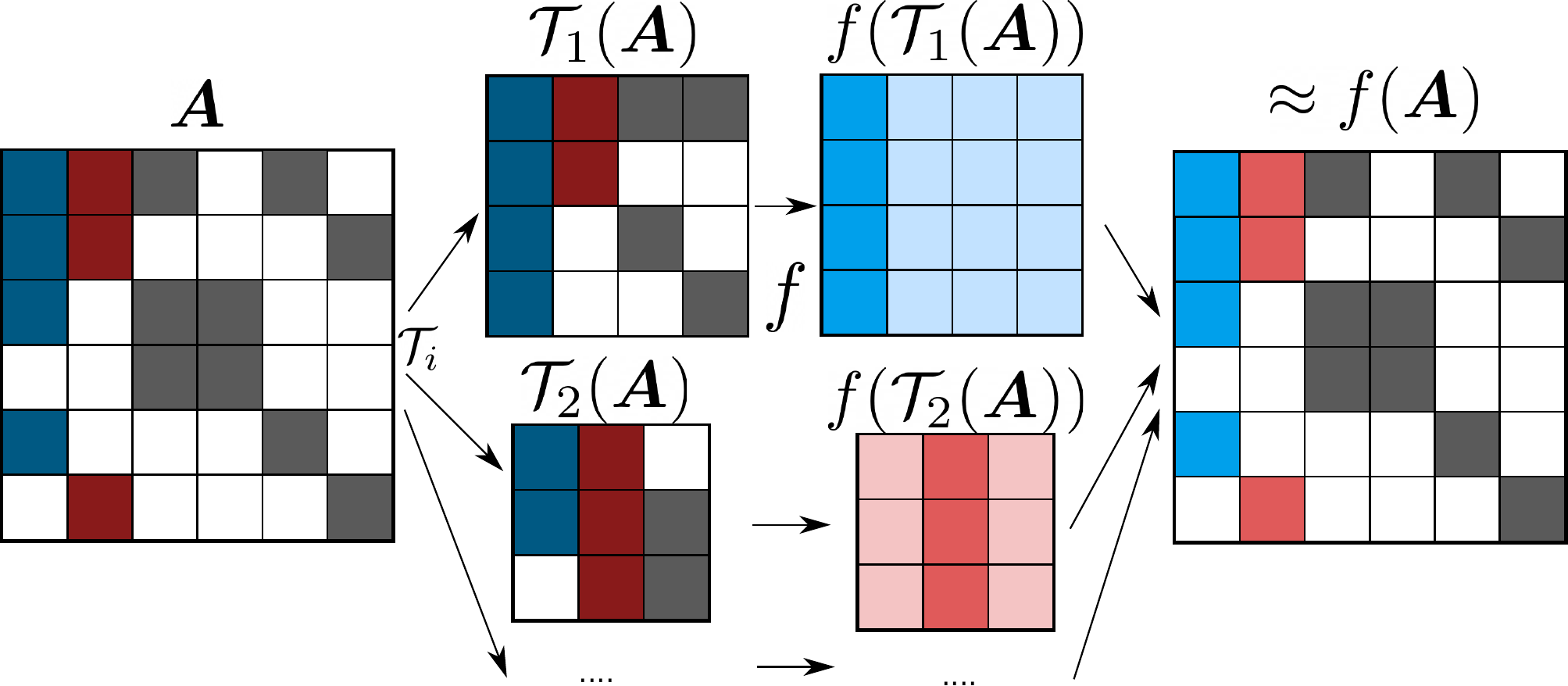}
    \caption{Schematic representation of the steps of the submatrix method for the approximate calculation of a matrix function $f(\mathbf{A})$ of a large sparse matrix $\mathbf{A}$. The first step is the construction of a submatrix $\mathcal{T}_i(\mathbf{A})$ for every column of the matrix $\mathbf{A}$. Then the matrix function is applied to the dense submatrices, i.e. $f(\mathcal{T}_i(\mathbf{A}))$, and finally the relevant result columns are inserted into the sparse result matrix. Figure reproduced from~\cite{SchadeDFT} under \href{http://creativecommons.org/licenses/by/4.0/}{CC-BY}.}
    \label{fig:submatrix}
\end{figure}

In recent work, we have developed the submatrix method that is specifically designed with these design principles in mind. The core idea, as illustrated in Fig.~\ref{fig:submatrix}, is to convert the evaluation of a matrix function on a large distributed sparse matrix into a large-scale parallel evaluation of the matrix function on many dense, but much smaller matrices. 
In combination with the purification scheme of Eq.~\ref{eq:purification}, 
we were able to perform record sized linear-scaling electronic structure computations on systems with more than 100 million atoms, thereby achieving a sustained performance of 324~PFLOPs with an efficiency of more than $67\%$ \cite{SchadeDFT}. 

\section{Concluding Remarks}

We conclude by noting that beside exploiting upcoming exascale supercomputers to facilitate ever more complex and accurate simulations, in the future alternative approaches such as deep neural networks (DNN), as well as quantum computing algorithms will find their application within CP2K. On the one hand this can be so-called learning on-the-fly schemes, in which the decision if the calculation of an observable can be outsourced to an already sufficiently accurate DNN or needs to be conducted explicitly, which can then be used as training data to directly relearn the DNN, is done during the simulation itself. On the other hand, all sort of hybrid approaches can be imagined, where the majority of a particular quantity is explicitly calculated using a computationally simple, but approximate electronic structure method, which is augmented by a correction term that is represented by a DNN for instance \cite{DeltaNN}. De facto exact configuration interaction, or reduced density-matrix functional (RDMF) theory simulations will become feasible by the usage of hybrid quantum-classical algorithms in which the quantum mechanical expectation values of the RDMF are evaluated on a quantum computer, whereas the parameters of the trial states are optimized by a Car-Parrinello-like constrained minimization scheme within CP2K on a classical computer \cite{SchadeRDMFT}.

\section{Acknowledgements}
The authors would like to thank the whole CP2K development team, who had been contributing to the code over the last two decades.

\end{document}